\documentclass[aps,prl,reprint]{revtex4-1}

\usepackage[draft]{todonotes}
\usepackage{xspace}
\usepackage{xcolor}

\newcommand{\figref}[1]{Fig. \ref{#1}}

\newcommand{\wcexclude}[1]{}

\renewcommand{\wcexclude}[1]{#1}

\newcommand{\PaperTitle}{Deep neural networks for direct, featureless learning through observation: the case of 2d spin models}

\let\explicitsection\section

\newcommand{\exclude}[1]{}

\begin{document}

\include{title}
\include{fillins}
\title{\PaperTitle}

\author{Kyle Mills}
\email[]{kyle.mills@uoit.net}
\affiliation{Dept. of Physics, University of Ontario Institute of Technology}

\author{Isaac Tamblyn}
\email[]{isaac.tamblyn@nrc.ca}
\affiliation{Dept. of Physics, University of Ontario Institute of Technology, University of Ottawa \& National Research Council of Canada}

\newcommand{\bold}[1]{#1}

\newcommand{\RSfourbyfouraccuracy}{99.88\%\xspace}
\newcommand{\TSsmalldatasetaccuracy}{98.98\%\xspace}

\newcommand{\TSbestperformingmodelnumberofexamples}{27,000\xspace}
\newcommand{\TSbestperformingmodelaccuracy}{100\%\xspace}
\newcommand{\TSNumseeds}{3\xspace}

\newcommand{\eightByEightBestNumberOfExamples}{100,000\xspace}
\newcommand{\eightByEightBestAccuracy}{99.922\%\xspace}
\newcommand{\eightClassificationAccuracy}{99.885\%\xspace}
\newcommand{\eightFrequencyWithinOne}{100\%\xspace}

\newcommand{\eightMAERegression}{1.782$J$\xspace}

\newcommand{\eightMAEMagnetization}{$2\times 10^{-3}$ per spin\xspace}
\newcommand{\fourMAEMagnetization}{$4\times 10^{-3}$ per spin\xspace}

\newcommand{\eightKRRMAE}{60.3$J$\xspace}
\newcommand{\eightRFMAE}{5.8$J$\xspace}

\newcommand{\yukawaMAE}{0.640$J$\xspace}
\newcommand{\yukawaSinMAE}{0.253$J$\xspace}
\newcommand{\pottsMAE}{\bold{0.542$J$\xspace}}

\newcommand{\longrangespeedup}{1600\xspace}
\newcommand{\SI}{Supplementary Information\xspace}

\date{\today}

\begin{abstract}
\wcexclude{\bold{
We demonstrate the capability of a convolutional deep neural network in predicting the nearest-neighbour energy of the $4\times 4$ Ising model.  Using its success at this task, we motivate the study of the larger $8\times 8$ Ising model, showing that the  deep neural network can learn the nearest-neighbour Ising Hamiltonian after only seeing a vanishingly small fraction of configuration space.  Additionally, we show that the neural network has learned both the energy and magnetization operators with sufficient accuracy to replicate the low-temperature Ising phase transition.  We then demonstrate the ability of the neural network to learn other spin models, teaching the convolutional deep neural network to accurately predict the long-range interaction of a screened Coulomb Hamiltonian, a sinusoidally attenuated screened Coulomb Hamiltonian, and a modified Potts model Hamiltonian.  In the case of the long-range interaction, we demonstrate the ability of the neural network to recover the phase transition with equivalent accuracy to the numerically exact method. Furthermore, in the case of the long-range interaction, the benefits of the neural network become apparent; it is able to make predictions with a high degree of accuracy, and do so \longrangespeedup times faster than a CUDA-optimized exact calculation. Additionally, we demonstrate how the neural network succeeds at these tasks by looking at the weights learned in a simplified demonstration. }
}
\end{abstract}


\wcexclude{
\maketitle
}

\section{Introduction}

The collective behaviour of interacting particles, whether electrons, atoms, or magnetic moments, is the core of condensed matter physics. The difficulties associated with modelling such systems arise due to the enormous number of free parameters defining a near-infinite configuration space for systems of many particles.  In these situations, where exact treatment is impossible, machine learning methods have been used to build better approximations and gain useful insight. This includes the areas of 
dynamical mean-field theory, strongly correlated materials, phase classification, and
materials exploration and design  \cite{Ovchinnikov2009,Kusne2014,Jesse2016,Balachandran2016,Carleo2017,Arsenault2014,Chng2016a,VanNieuwenburg2016}.

As an introductory many-particle system, one is commonly presented with the square two-dimensional Ising model, a ubiquitous example of a ferromagnetic system of particles.  The model consists of an $L\times L$ grid of discrete interacting ``particles'' which either possess a spin up ($\sigma = 1$) or spin down ($\sigma = -1$) moment.  The internal energy associated with a given configuration of spins is given by the Hamiltonian $\hat H = -J \sum\sigma_i\sigma_j  $
where the sum is computed over all nearest-neighbour pairs ($\langle i,j\rangle$), and $J$ is the interaction strength. For \bold{$J>0$}, the system behaves ferromagnetically; there is an energetic cost of having opposing neighbouring spins, and neighbouring aligned spins are energetically favourable.

The canonical Ising model defined on an infinite domain (i.e. periodic boundary conditions) is an example of a simple system which exhibits a well-understood continuous phase transition at a critical temperature $T_c \approx 2.269$.
At temperatures below the critical temperature,
the system exhibits highly ordered behaviour, with most of the spins in the system aligned.  Above the critical temperature,
the system exhibits disorder, with, on average roughly equivalent numbers of spin up and spin down particles.  The ``disorder'' in the system can be represented by an order parameter known as the ``magnetization'' $M$, which is merely the average of all $L^2$ individual spins.

Configurations of the Ising model can be thought to belong to one of two phases. Artificial neural networks have been shown to differentiate between the phases \cite{Carrasquilla2016,Wang2016a},  effectively discovering phase transitions.  This is, however, merely a binary classification problem based on the magnetization order parameter. The membership of a configuration to either the high- or low-energy class does not depend upon any interaction between particles within the configuration.  Furthermore, convolutional neural networks have been trained to estimate critical parameters of Ising systems \cite{Tanaka2016}.  Machine learning methods have been demonstrated previously in many-body physics applications \cite{Arsenault2014} and other two-dimensional topological models are discussed frequently \cite{Kitaev2006,Levin2006} in quantum field theory research. However, the use of deep convolutional neural networks remains infrequent, despite their recently presented parallels to renormalization group \cite{Mehta2014} and their frequent successes in difficult machine learning and computer vision tasks, some occurring a decade ahead of expectations \cite{Mnih2013,Silver2016}.  

We demonstrate that a convolutional deep neural network, trained on a sufficient number of examples, can take the place of conventional operators \bold{for a variety of spin models}.  Traditional machine learning techniques depend upon the selection of a set of descriptors (features) \cite{Ghiringhelli2015}. Convolutional deep neural networks have the ability to establish a set of relevant features without human intervention, by exploiting the spatial structure of input data (e.g. images, arrays). Without human bias, they detect and optimize a set of features, and ultimately map this feature set to a quantity of interest.  For this reason, we choose to call deep convolutional neural networks ``featureless learning''.  \bold{We take a more in-depth look into this process in the section titled ``A closer look at the convolutional kernels'' later in this work.}  Furthermore, we demonstrate that a neural network trained in this way can be practically applied in a simulation to accurately compute the temperature dependence of the heat capacity.  In doing so, we observe its peak at the critical temperature, a well-understood \cite{Ferdinand1969} indication of the Ising phase transition. \bold{Additionally, we investigate the effectiveness of the deep learning approach on three additional spin model Hamiltonians.}

\section{Methods}

\subsection{Deep learning}

\begin{figure}
\begin{center}
 \includegraphics[width=0.99\columnwidth]{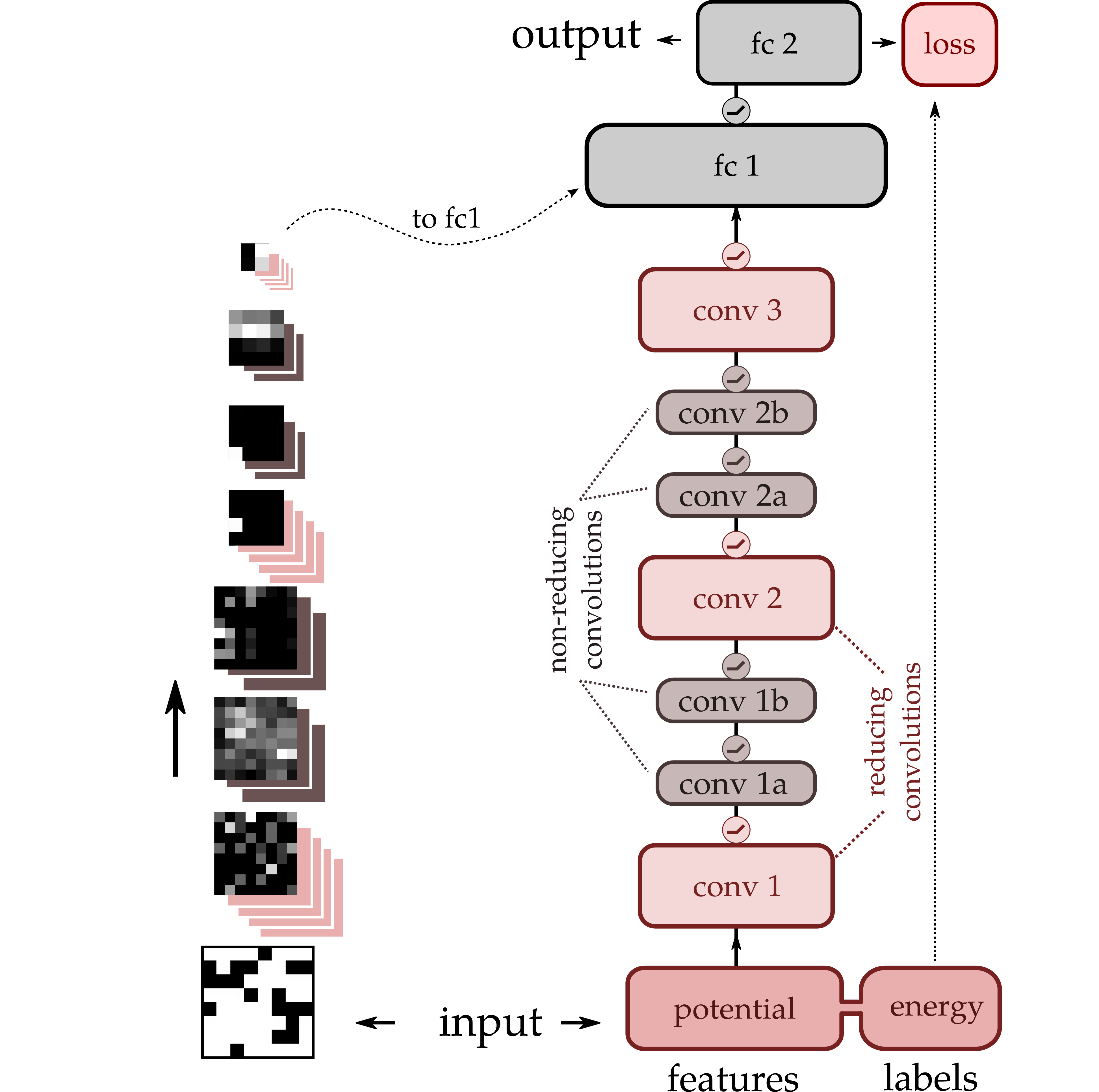}
 \caption{ Deep neural network architecture used for predicting Ising model operators.  The network consists of 7 convolutional layers with two fully-connected layers at the top. On the left, the output of a given filter path is shown for an example Ising configuration. The final $2\times 2$ output is fed into a wide fully-connected layer. ReLU (recified linear unit) activation is used on all convolutional layers. \label{isingdeepnetschematic}}
 \end{center}
\end{figure}

We used a very simple deep neural network architecture, shown in \figref{isingdeepnetschematic}.  In previous work, we demonstrated that the same neural network structure, differing only in the number of repeat units, was capable of predicting quantities associated with the one-electron Schr\"{o}dinger equation \cite{Mills2017}.  In this network, the convolutional layers
work by successively mapping an image into a feature space where interpolation is possible using a nonlinear boundary.  The final decision layers 
impose this boundary and produce an output value.   
Other \bold{methods can be trained in a featureless learning fashion,} such as kernel ridge regression and random forests. We compare our approach to these methods in ``DNN versus other methods'' below.

Our neural network consists of 7 subsequent convolutional layers.  We use two different types of convolutional layers, which we call ``reducing'' and ``non-reducing''.

The 3 reducing layers operate with filter (kernel) sizes of $3\times 3$ pixels.  Each reducing layer operates with 32 filters and a stride of $2\times 2$, effectively reducing the data resolution by a factor of two at each step.  In between each pair of these reducing convolutional layers, we have inserted two convolutional layers (for a total of 4) which operate with $16$ filters of size $4\times4$. These filters have unit stride, and therefore preserve the resolution of the image.  The purpose of these layers is to add additional trainable parameters to the network, and we have previously \cite{Mills2017} found that 2 was a good balance between speed and accuracy.  All convolutional layers have ReLU (rectified linear unit) activation.

The final convolutional layer is fed into a fully-connected layer of width 1024, also with ReLU activation. This layer feeds into a final fully-connected layer of size 15.  This output can be interpreted as a vector of the probability of membership to each energy class. For the larger $8\times 8$ configurations, there are 63 possible energy values, and therefore the final fully-connected layer is modified to have a width of 63.  This output is used to compute the softmax cross-entropy loss. 

To train the models, we minimized this loss function using the AdaDelta \cite{Zeiler2012} optimization scheme, with a global learning rate of 0.001.  We monitored the loss function as training proceeded and terminated training after the loss function appeared to converge sufficiently. 

Training of the models was carried out on a custom-built computer housing multiple  graphical processing units. We used a custom TensorFlow \cite{GoogleResearch2015} implementation in order to make use of the GPUs in parallel.  We placed a complete copy of the neural network on each GPU, so that each can compute a forward and back-propagation iteration on one full batch of images.  Our effective batch size was 800 images per iteration.  After each iteration, the GPUs share their independently computed gradients and the optimizer proceeds to adjust the parameters in the direction that minimizes the loss function.

The series of convolutional layers in this network is designed to operate on images of size $16\times 16$.
For this reason, when training the $4\times 4$ Ising model, we perform lossless upscaling by repeating each row and column 4 times to achieve a compatible input size.  With the $8\times8$ configurations, we upscale by a factor of 2 to obtain the same input size.  This does not notably affect the performance of the models, and it permits the use of the same network architecture for both sizes.  In practice, one could use a layout similar to that suggested in Ref. \cite{Luchak2017} to accommodate arbitrarily-sized Ising model grids, including grids differing in size to those in the training set.

\section{Datasets}

\begin{figure}
 \includegraphics[width=0.99\columnwidth]{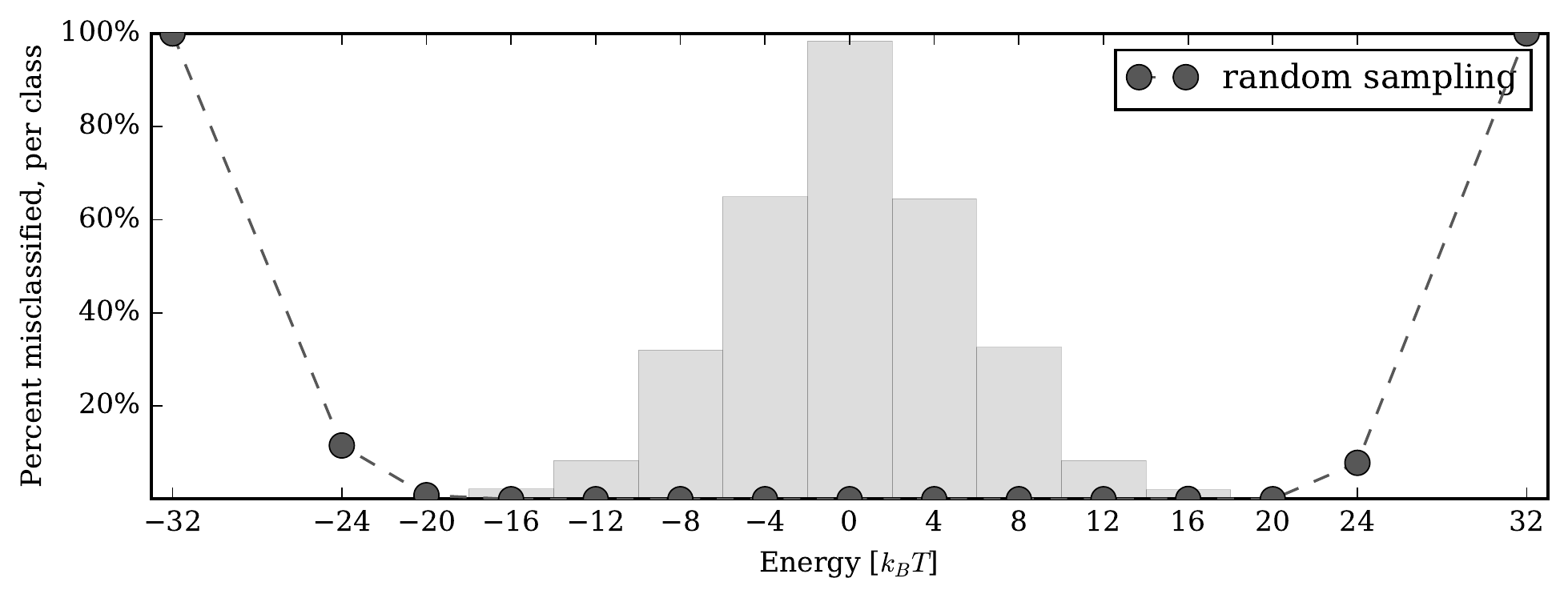}
 \caption[$4\times 4$ Ising accuracy, random sampling]{The overall accuracy for the model trained on randomly sampled data is \RSfourbyfouraccuracy, however the neural network misclassifies the majority of high- and low-energy configurations.  With larger Ising models, this effect would be more significant as the central degeneracy is greater. \label{4x4_RS_accuracy_vs_class}}
\end{figure}

Two dimensional \bold{spin} model configurations \bold{(with the exception of the Potts model)} can be respresented as binary-valued arrays, with each element having a value of either $\sigma = -1$ (spin down) or $\sigma=1$ (spin up).  As such, a simple method to generate an arbitrarily large amount of training data is to randomly draw the state of each spin, with uniform probability of it being $-1$ and $1$.  This method, while trivially easy to implement, results in an energy distribution centered sharply around zero (histogram of \figref{4x4_RS_accuracy_vs_class}), since the central energy levels of the Ising model are highly degenerate.  There is very little probability of generating a high-energy (``checkerboard'') or low-energy (``solid'') configuration.  This will be problematic to the application of the deep learning model, as the neural network will not have been exposed to features present in the high and low-energy configurations during training.  We initially trained the neural network naively on approximately 12500 randomly generated training examples. The training dataset contains only 10343 of the 65536 possible configurations (16\%).

We evaluated this model on the complete set of 65536 $4\times4$ configurations and it achieved an accuracy of \RSfourbyfouraccuracy (\RSfourbyfouraccuracy of the configurations were classified correctly). While this appears to be excellent performance, closer inspection reveals that many configurations with energies below $-16J$, and above $16J$, are misclassified, as shown in \figref{4x4_RS_accuracy_vs_class}.  This problem would greatly affect a Monte Carlo simulation replicating the low-temperature phase transition, as this phenomenon is dependent upon the correct evaluation of low-energy configurations. 

\begin{figure}
 \includegraphics[width=0.99\columnwidth]{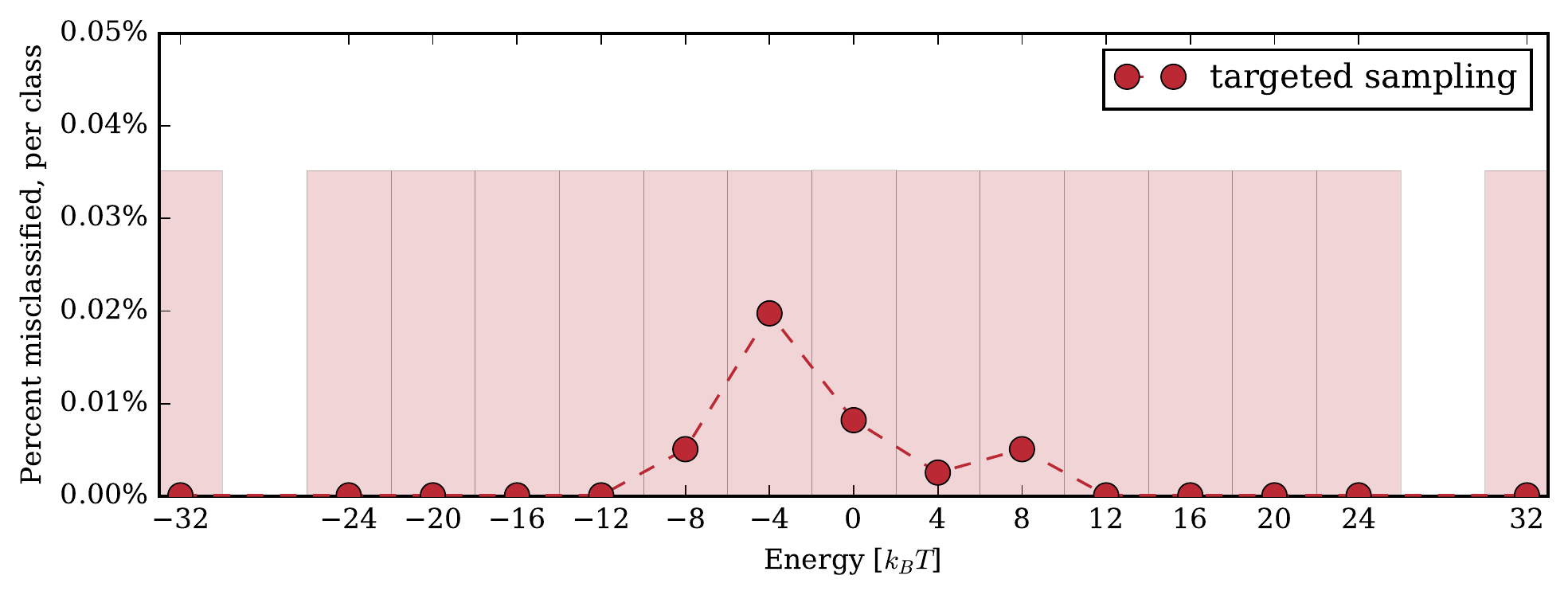}
 \caption[]{When trained on an even distribution of energies (targeted sampling dataset) the deep neural network was able to classify all but a handful of configurations. The misclassified configurations appear central to the energy distribution as there is more variation in this region.\label{4x4_TS_accuracy_vs_class}  }
\end{figure}

The motivation for a more intelligent sampling method is clear; a more even distribution of energies is necessary if one wishes to accurately predict the low- and high-energy configurations.  We implemented a modified form of umbrella sampling, which we have named ``targeted sampling" (TS) in an attempt to achieve a more even distribution of energies.  This sampling method resembles the Metropolis-Hastings algorithm in structure, but instead of seeking low-energy states, we ``target'' specific energies, accepting configurations that move us toward the target energy, and rejecting ones (with a Gaussian probability) that lead us away.  In this way, we can collect examples across the energy spectrum in an intelligent way, achieving a very even distribution of energies, as seen in the histogram of \figref{4x4_TS_accuracy_vs_class}.

\section{Results}

\subsection{The $4\times4$ Ising model}

We begin our investigation with the $4\times4$ Ising model, as the configuration space is of a manageable size that we can easily compute \textit{all possible configurations} ($65536$ total unique configurations).  
Because of this, we can explicitly evaluate how well a model performs by evaluating the model on the entirety of configuration space.  
The energies of these configurations are discrete, taking on 15 possible values (for the $4\times 4$ model) ranging from $-32J$ to $32J$.  The discrete energy values allow us to treat energy prediction as a machine learning ``classification problem''. 
In the areas of handwriting recognition and image classification, deep convolutional neural networks with such an output structure have excelled time and time again \cite{Lecun1998,Simard2003,ciresan2011flexibles,Szegedy2014}. The value of $J$ can be any constant, as the input and output of the neural network can be scaled appropriately.  In this work we use $J=1$.

\begin{figure}
 \includegraphics[width=0.99\columnwidth]{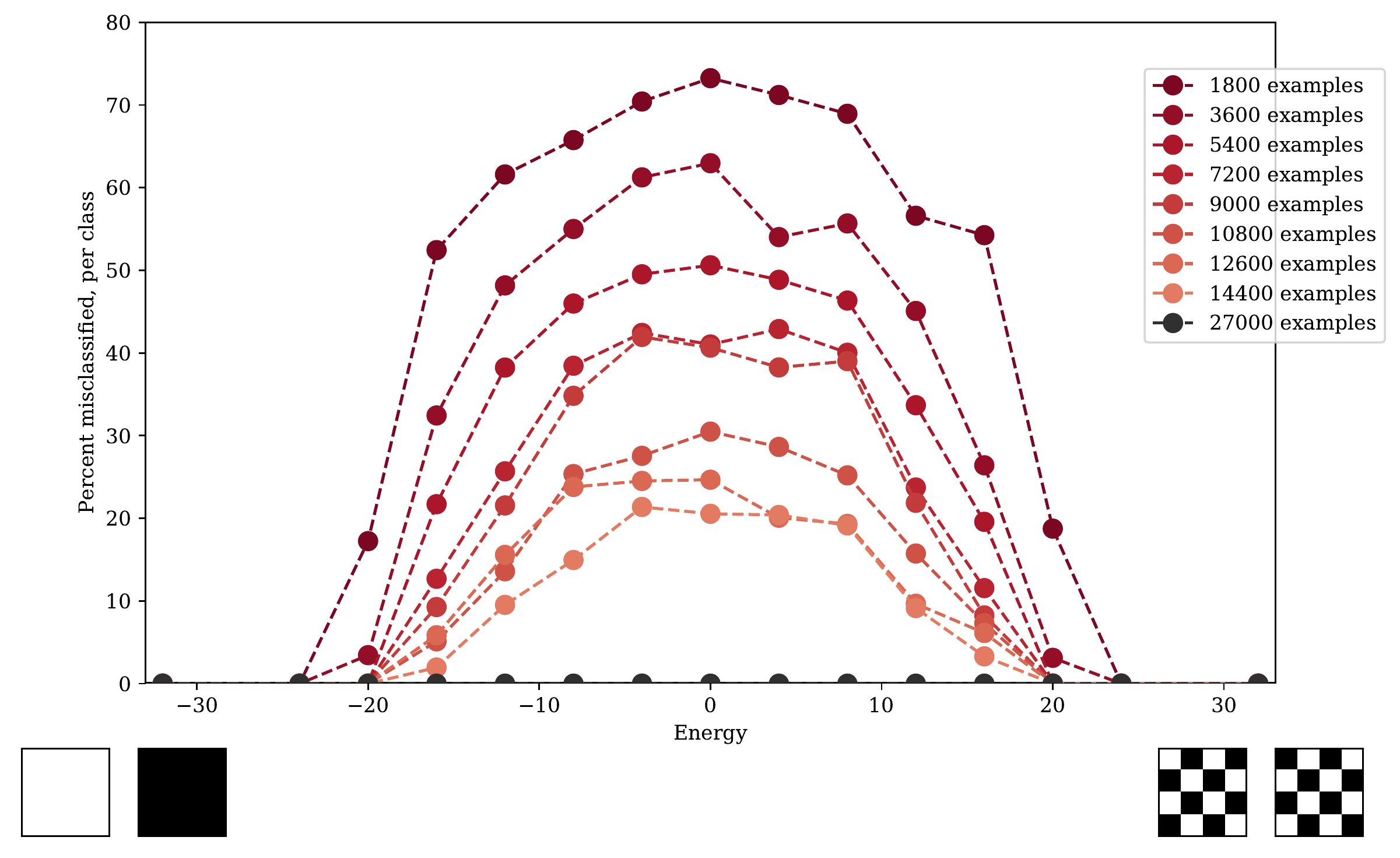}
 \caption{We investigate how the classification accuracy of the neural network depends on the number of training examples.  Since 27000 training examples resulted in almost perfect accuracy, we chose it as the number, giving rise to 1800 configurations per class. \label{ErrorVsClassVsNumerofExamples}  }
\end{figure}

\begin{figure}
 \includegraphics[width=0.75\columnwidth]{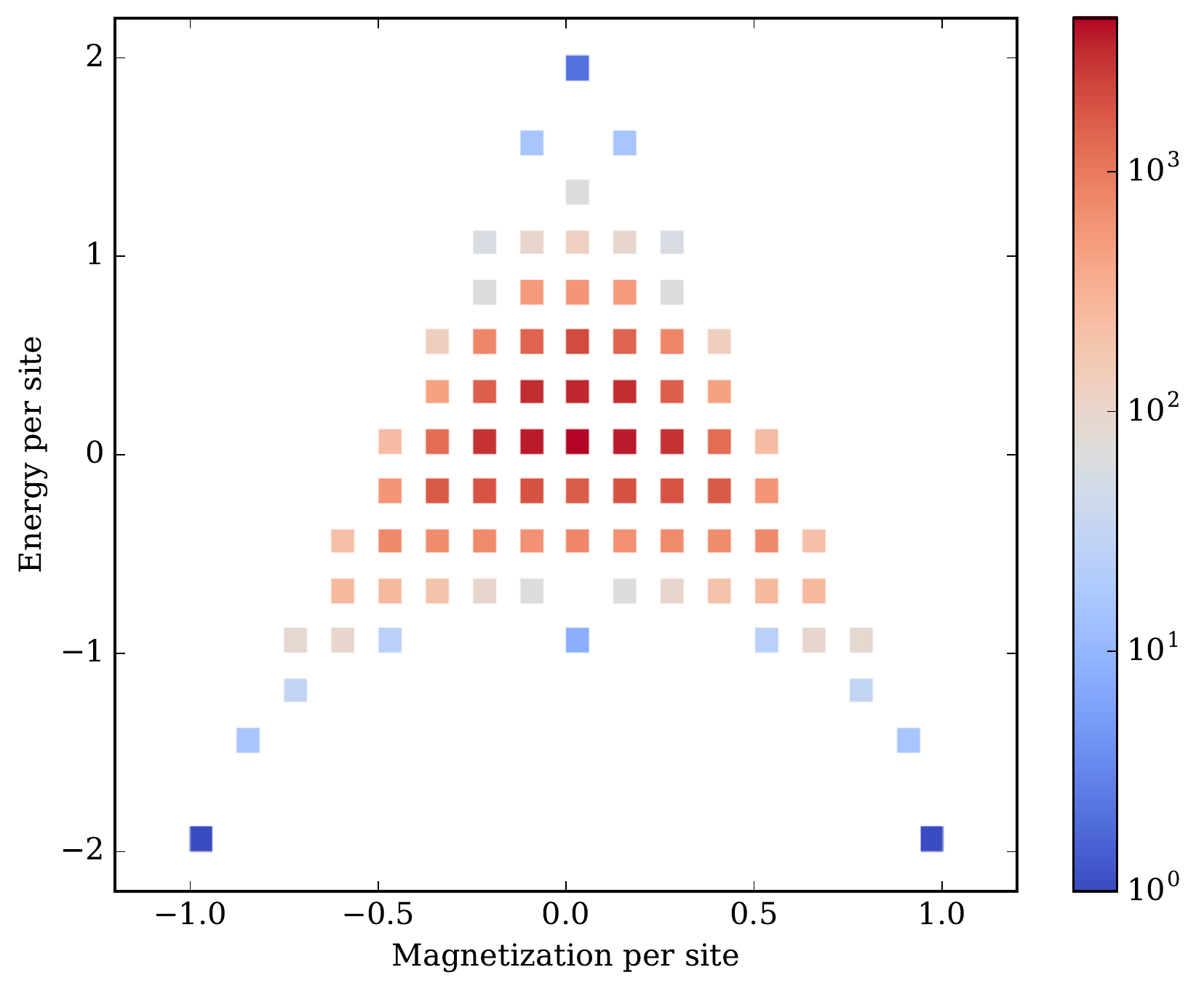}
 \caption{Configuration space of the $4\times 4$ Ising model. The colors represent the density of states.\label{fig:isingconfigurationspace}  }
\end{figure}

We generated three independent datasets, each consisting of 27,000 training examples (1800 examples per class),  gathered using the targeted sampling approach.
In \figref{ErrorVsClassVsNumerofExamples}, we show the classification accuracy of the deep neural network with respect to the number of example configurations provided during training.  Exceptional accuracy is achieved at 27000 training examples (1800 per class). Each dataset contains less than 20\% of configuration space (some energy classes are over-sampled to fill the 1800-example quota).  We trained our neural network architecture on each of these three datasets. The neural network was able to classify all but a handful of Ising configurations, on average. On one dataset, it achieved an accuracy of \TSbestperformingmodelaccuracy. In all cases of misclassification, 100\% of misclassified examples only have an error of $\pm 1$ energy level, indicating the neural network is just barely failing to classify such examples. All misclassified configurations had energies near zero. In this region there is considerable variation due to the degeneracy of the Ising model (apparent in \figref{fig:isingconfigurationspace}), and therefore predictions based on a uniform number of training examples per class are slightly more challenging. At the extreme energies ($\pm 32$), individual configurations are repeated many times in order to fill the quota of training examples. It is worth noting again that this neural network had access to less than 20\% of configuration space so it is clearly correctly inferring information about examples it has not yet seen.

\subsubsection{The $8\times 8$ Ising model}

Although the $4\times 4$ model is instructive, larger Ising models such as the $8\times 8$ model are interesting since the enormity of configuration space \mbox{($2^{8^2}\approx 10^{19}$)} precludes training on even a modest fraction of possible configurations, so the neural network truly needs to ``learn'' how to predict energies from seeing only a minuscule fraction of configuration space.

\begin{figure}
\begin{center}
 \includegraphics[width=0.99\columnwidth]{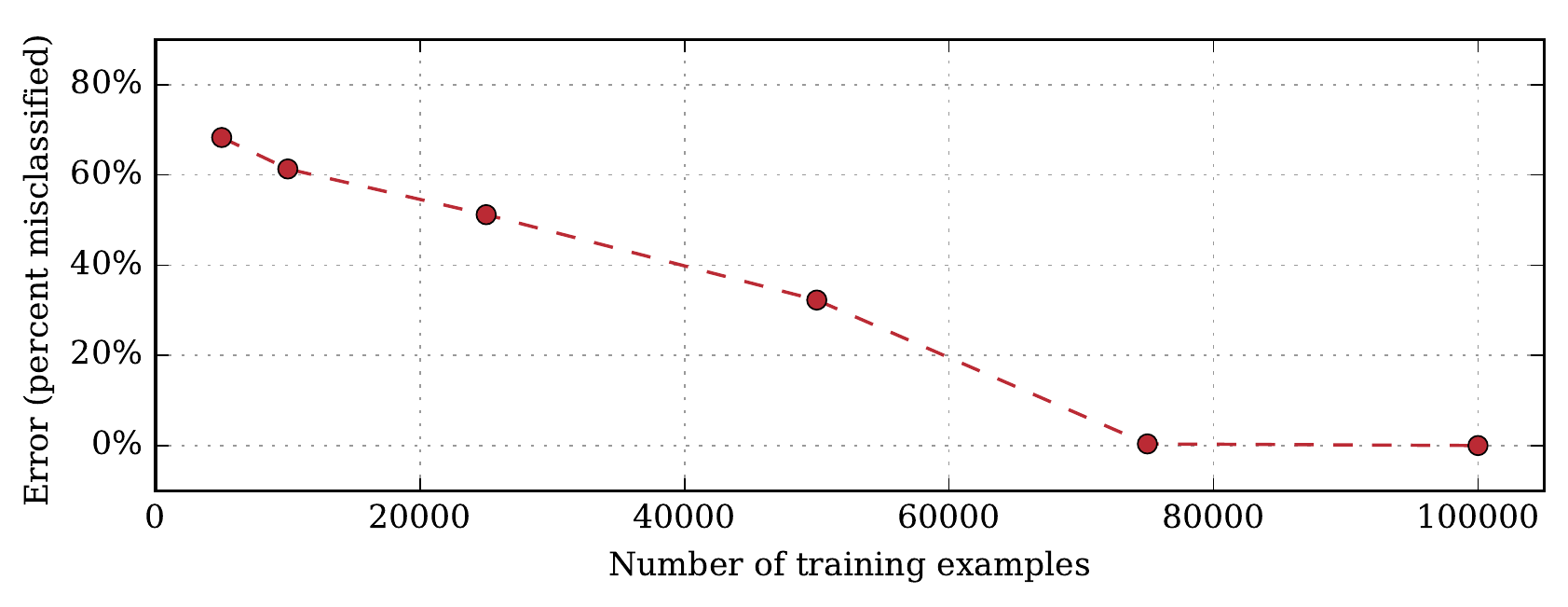}
 \caption{More training examples unsurprisingly leads to better performance.  At \eightByEightBestNumberOfExamples, the neural network is able to classify $8\times 8$ Ising configurations into their 63 discrete energy levels with \eightByEightBestAccuracy accuracy.  This represents a vanishly small subset of configuration space, the entirety of which consists of $2^{64}$ configurations. \label{8x8_accuracy_vs_examples}}
 \end{center}
\end{figure}

We performed a convergence study to determine the optimal number training examples.  At \eightByEightBestNumberOfExamples, the neural network is able to classify $8\times 8$ Ising configurations into their 63 discrete energy levels with \eightByEightBestAccuracy accuracy as shown in \figref{8x8_accuracy_vs_examples}. Note that we can no longer report an accuracy computed over the entirety of configuration space, so we must report the accuracy of the model on a separate testing set of data.

The testing dataset consists of 50,000 examples separated from the training dataset prior to training. No examples in the testing set appear in the training set. \bold{This is ensured by separating examples into testing and training based on their SHA256 hash}.  Importantly, as with the $4\times4$ model, in the few cases where the model did fail, it did not fail by very much: \eightFrequencyWithinOne of the time, the predicted class is either correct or only one energy class away from correct.  The neural network does exceptionally well at predicting energies when only exposed to a small subset of configuration space during training.

\subsubsection{Regression}
In practice, a deep neural network capable of classifying configurations into well-defined bins is less appealing than one which could predict continuous variables.  ``Real-life'' systems rarely exhibit observables and characteristics that are quantized at the scales relevant to the macroscopic problem.  As such, this is a good opportunity to investigate a form of deep neural network output structure known as ``regression''.  In a regression network, instead of the final fully connected layer having a width equal to the number of classes, we use a fully connected layer of width 1: a single output value. In this case, a softmax cross-entropy loss layer is no longer appropriate.  The simplest form of loss function for a single-output regression network is the mean-squared error between network predictions and the true value of the energy.

We modify the deep neural network in this way to perform regression.  Changing nothing about the training process other than the loss function, we see that the model performs quite well, with a median absolute error of \eightMAERegression. With the Ising model, the allowed energy classes are separated by 4 energy units, so an error of \eightMAERegression is consistent with the capability of the network to accurately classify examples into these bins of width 4. Additionally, we trained the deep neural network to learn the magnetization; it performs exceptionally well with a median absolute error of \fourMAEMagnetization  and \eightMAEMagnetization for the $4\times4$ and $8\times 8$ models, respectively.  This is not particularly surprising as the magnetization is a very simple, non-interacting operator.  This effectively amounts to using a convolutional neural network to compute the sum of an array; we present it merely as a demonstration of a neural network's ability to learn multiple mappings.

\subsubsection{Replicating the Ising Model phase transition}

The Ising model defined on an infinite domain exhibits a phase transition at a critical temperature $T_c \approx 2.269$.
For a finite domain under periodic boundary conditions, however, a correction factor $\gamma$ is necessary to compensate for the correlations between periodic lattice images. This behaviour is discussed in detail in Ferdinand and Fisher's 1969 work (ref. \citep{Ferdinand1969}), and in this analysis we will denote the ``theoretical'' critical temperature as $\gamma T_c$.

Using a Metropolis-Hastings algorithm, one can sample the configuration space accessible to the Ising model at a given temperature.  Using a Boltzmann rejection probability, the mean energy per site, $\bar E$ can be computed for a given temperature.  Repeating for various temperatures allows one to plot $\bar E$ against $T$ and observe the phase transition.  The Metropolis-Hastings algorithm, and thus the demonstration of this phase transition, depends on the accurate evaluation of Ising configuration energies. \bold{As with any mathematical model, the ultimate test is its ability to make predictions of sufficient quality that one can recover a physically realistic phenomena.}

\begin{figure}
 \includegraphics[width=0.99\columnwidth]{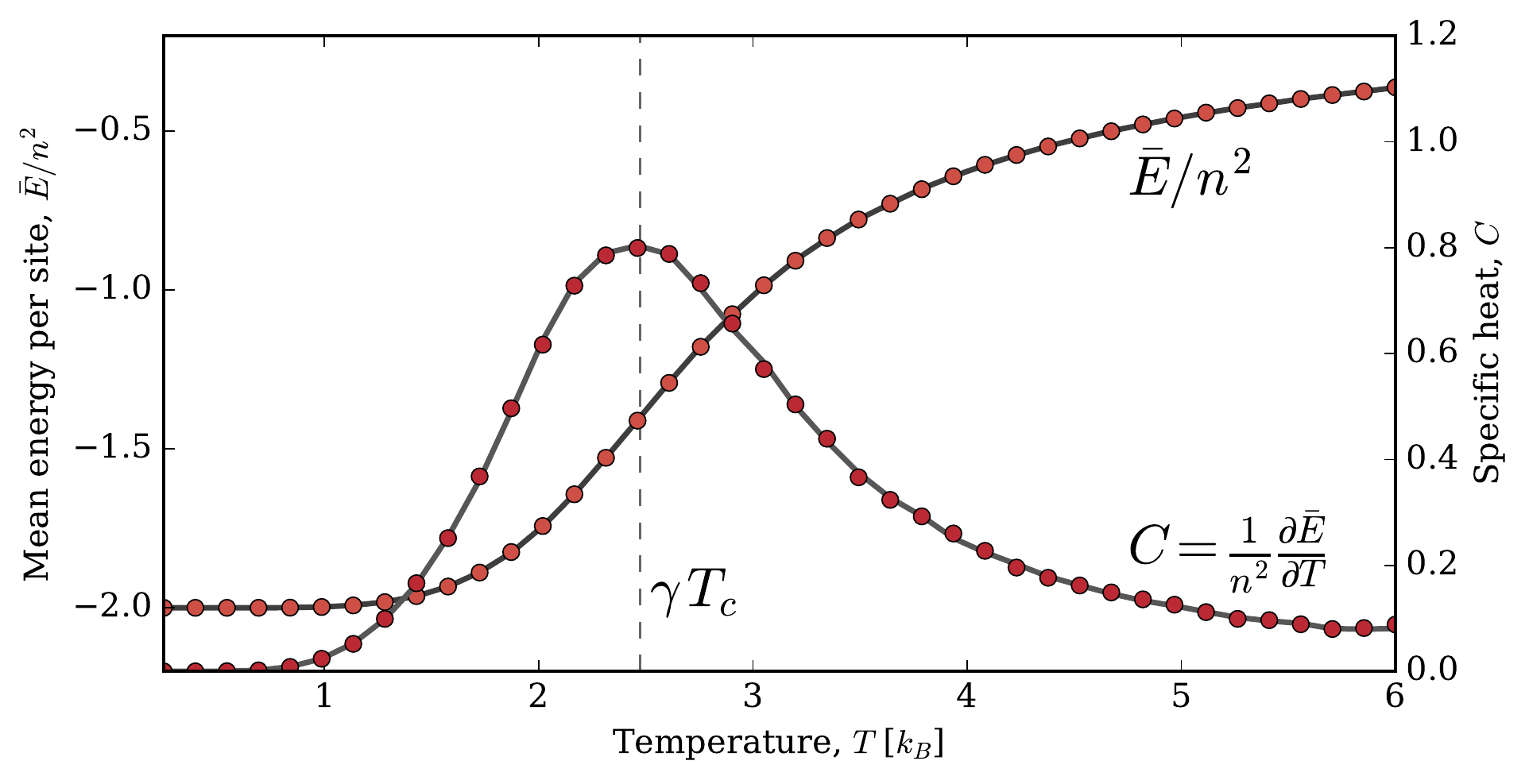}
\caption[$4\times 4$ phase diagram]{The average energy per site at various temperatures, as well as the heat capacity, $C$ for the $4\times 4$ Ising model.  The solid lines and dots indicate the energy evaluation methods used: the exact Hamiltonian, and DNN, respectively. These results are averaged over 400 independent simulations. The standard deviation is negligibly small. \label{fig:phaseplot4x4}}
\end{figure}

\begin{figure}
 \includegraphics[width=0.99\columnwidth]{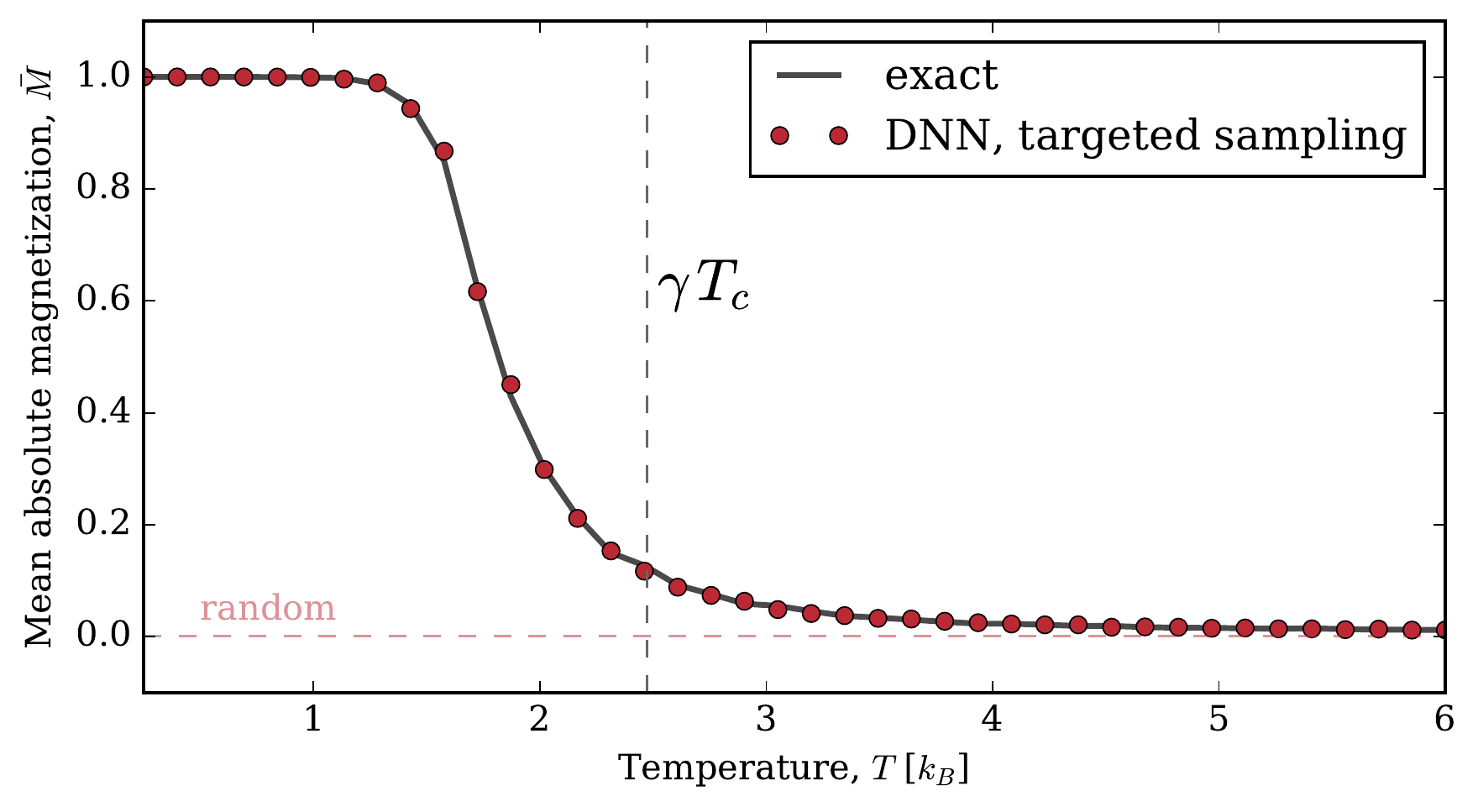}
\caption{The magnetization per site at various temperatures for the $4\times 4$ Ising model.  The solid line denotes the simulation using the exact magnetization and energy operators, and the dots represent the deep-learned energy and magnetization operators. These results are averaged over 400 independent simulations. The standard deviation is negligibly small. The horizontal dashed line indicates the mean absolute magnetization of a purely random distribution of spins (the entropy-dominating high-temperature limit), which is close to, but not equal to zero. \label{fig:phaseMT4x4}}
\end{figure}

We generated the phase diagram for the $4\times 4$ Ising model, evaluating the energy using the exact Hamiltonian.
Then, we replaced the magnetization and energy operators with the trained deep neural networks.  The phase diagrams match exactly, and are presented in Figs. \ref{fig:phaseplot4x4} and \ref{fig:phaseMT4x4}.

\begin{figure}
 \includegraphics[width=0.99\columnwidth]{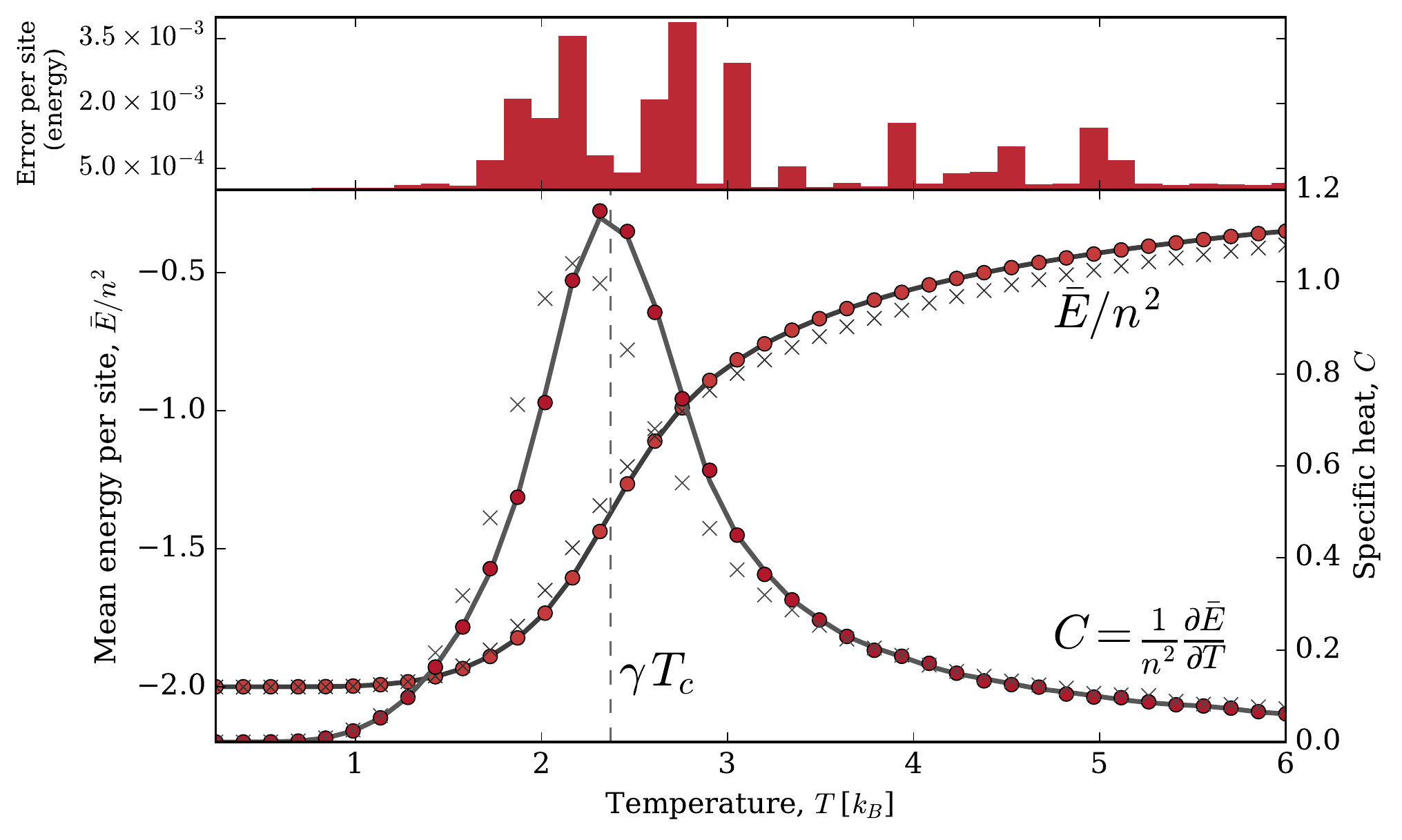}
\caption[$8\times 8$ phase diagram]{The average energy per site at various temperatures, as well as the heat capacity for the $8\times 8$ Ising model.  The solid line denotes the simulation using the analytic Hamiltonian, the dots represent the deep-learned Hamiltonian with classification, \bold{and the crosses represent the deep-learned Hamiltonian with regression. The absolute difference between the per-site-energies obtained from the analytic Hamiltonian and deep-learned classification model are plotted above.} These results are averaged over 400 independent simulations. The standard deviation is negligibly small. \label{fig:phaseplot8x8}}
\end{figure}

We repeated this exercise with \bold{both the $8\times 8$ classification and regression models, and observe the phase transition. As \figref{fig:phaseplot8x8} shows, in the case of classification, the deep neural network is able to learn the energy and magnetization operators with sufficient precision to replicate the phase transition.  In the case of regression, the phase transition is still observed, however at a slightly lower temperature.  This is not completely surprising, as the classification method effectively snaps any slightly incorrect predictions to the nearest correct value.
}

\subsubsection{Long-range interactions}
\begin{figure}
 \includegraphics[width=0.99\columnwidth]{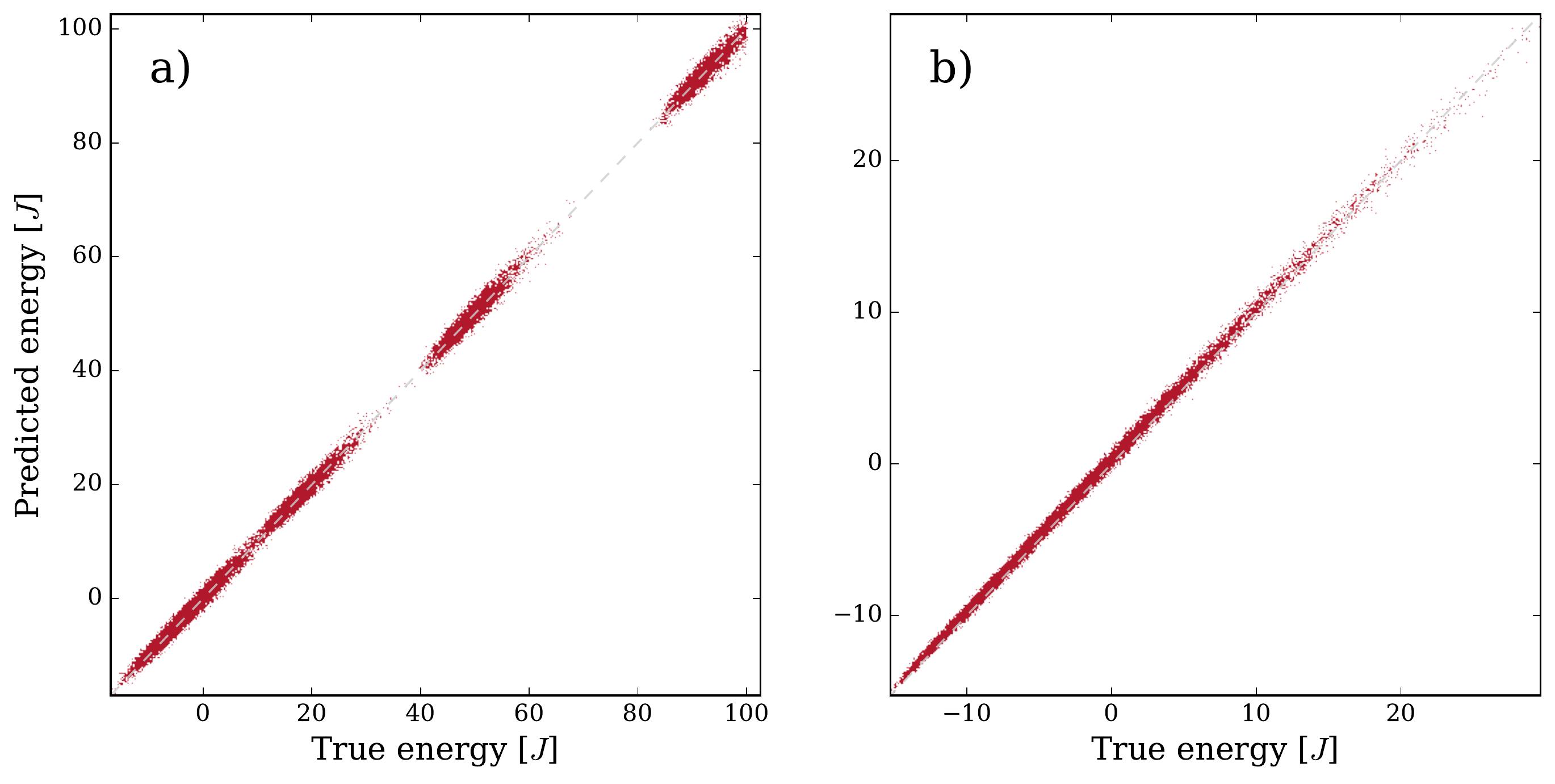}
\caption{The deep neural network is able to learn arbitrary long-range interactions with high accuracy.  Here we plot the DNN-predicted a) screened Coulomb energy, and b) sinusoidal screened Coulomb energy against the explicitly calculated energy for the 20,000 examples in the test set. The training and test set are randomly sampled.  \label{fig:yukawaPerf}}
\end{figure}

The extent of the traditional Ising Hamiltonian is very short-range, including only nearest-neighbour interactions. Physical systems frequently depend on long-range interations.  Herein, we demonstrate that the same deep neural network is able to learn the energies associated with two long-range interactions.  First, we demonstrate the screened Coulomb Hamiltonian: the traditional pairwise Coulomb interation attenuated by an exponential term \cite{Debye1923,Yukawa1934}.  We computed this energy for 120,000 $8\times 8$ Ising configurations using an explicit sum method and periodic boundary conditions, ensuring the infinite summation was converged sufficiently for the effective cutoff we used of 64 units, i.e. 8 times the size of the unit lattice.  The summation is very computationally expensive, as it must be computed for every pair of spins between the unit lattice and all periodic images until the effective cutoff radius is reached and the sum converges. Since the algorithm is amenable to parallelization, we implemented it in CUDA for performance \cite{Lam2015}.
We trained our deep neural network \bold{to perform regression} on a set of 100,000 examples, and tested the network on a non-intersecting set of 20,000 examples.  Our neural network is able to learn this long-range Hamiltonian with considerable accuracy, performing with a median absolute error of \yukawaMAE energy units.  The performance of the model is shown in \figref{fig:yukawaPerf}a.  

\begin{figure}
 \includegraphics[width=0.99\columnwidth]{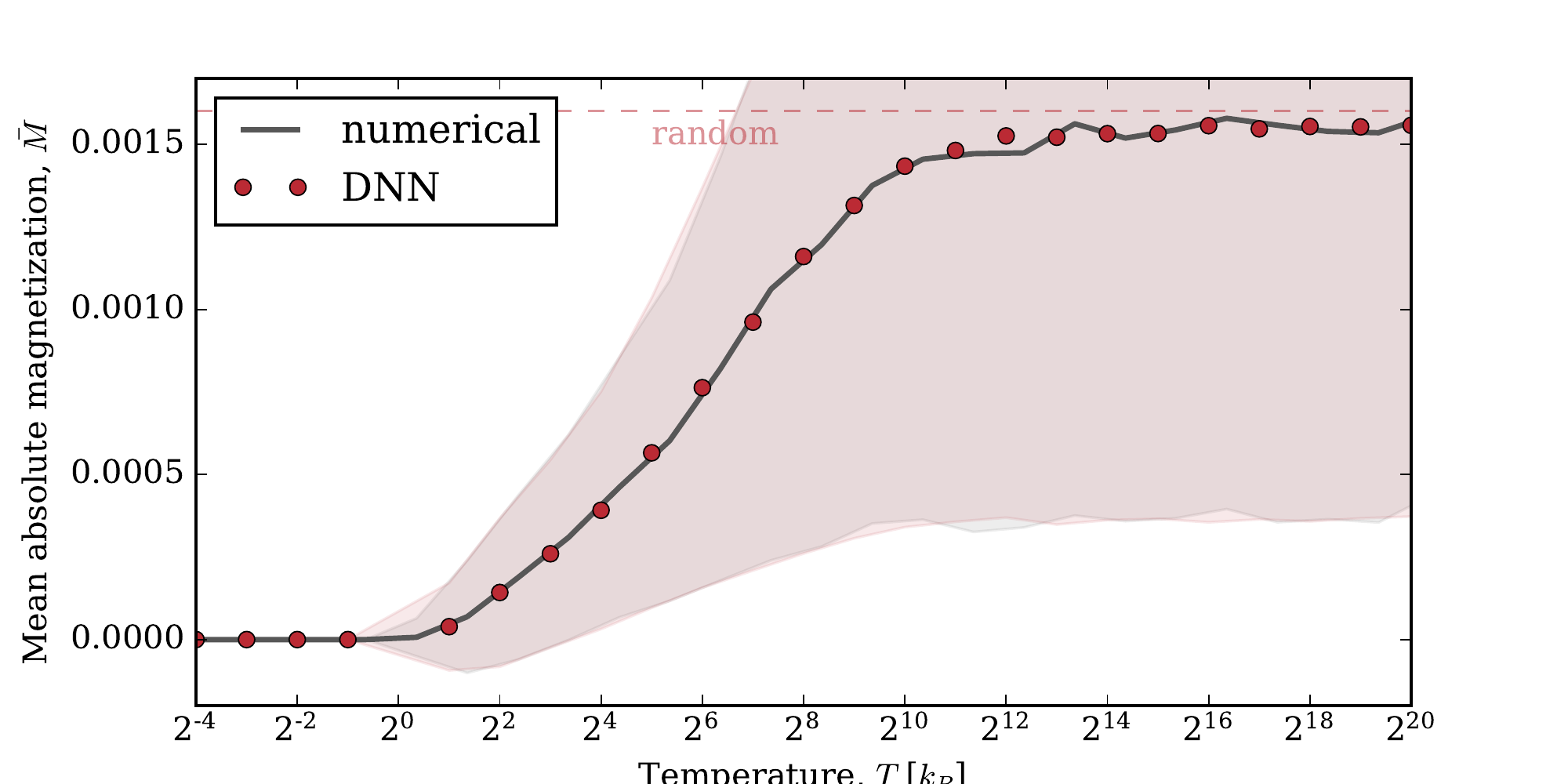}
\caption{\bold{The average magnetization per site at various temperatures for the (antiferromagnetic) long range screened Coulomb Hamiltonian. The solid line denotes the simulation using the numerical energy operator, and the dots represent the simulation run with the deep-learned energy operator. The dashed line labelled ``random'' denotes the mean absolute spin of a purely random distribution of spins, the entropy-dominating theoretical limit for infinite temperature. The filled area represents one standard deviation of the mean.}\label{fig:longrangephase}}
\end{figure}

\bold{Furthermore, we repeated the Metropolis-Hastings simulation and discovered a phase transition as the temperature increases. We then tuned our neural network to this ``thermally sampled'' data, and repeated the simulation using the The results are plotted in \figref{fig:longrangephase}.  Since this Hamiltonian represents an antiferromagnetic interaction, the most energetically stable configuration is the ``checkerboard'' configuration, in contrast to the ferromagnetic Ising model.  At (very) high temperature, the mean absolute magnetization approaches a value consistent with a purely randomly-drawn set of spins. Again our model performs well, matching the numerical simulation well.}

Secondly, to demonstrate the applicability of such a deep neural network architecture to arbitrary long range interactions, we modified the screened Coulomb Hamiltonian to have a sinusoidal dependence in $r$, e.g.
\begin{equation}
\hat H = J \sum\limits_{\{i,j\}}^{r_{\mathrm{cut}}}\sigma_i\sigma_j \frac{e^{-kr_{ij}}}{r_{ij}} \sin(r_{ij}),
\end{equation}
where the summation, like in the screened Coulomb Hamiltonian, is carried out over all combinations of spins between the configuration and all neighouring periodic images out to a radius of $r_{\mathrm{cut}}$.  This is, intentionally, a completely \textit{arbitrary} modification to the Hamiltonian made to demonstrate the wide generalizability of the deep neural network approach.  Following an identical training procedure as the screened Coulomb Hamiltonian, the deep neural network was able to make predictions with an accuracy of \yukawaSinMAE energy units.  The performance is plotted in \figref{fig:yukawaPerf}b.

While the accurate learning of the long-range interactions is in itself impressive, additionally the deep neural network drastically outperforms the explicit calculations in terms of speed.  The deep neural network can make predictions at a rate \longrangespeedup times faster than the CUDA-enabled ``exact'' calculation (performing at comparable median error), when running on a single NVIDIA Tesla K40.

\subsubsection{A modified Potts model}
\begin{figure}
 \includegraphics[width=0.58\columnwidth]{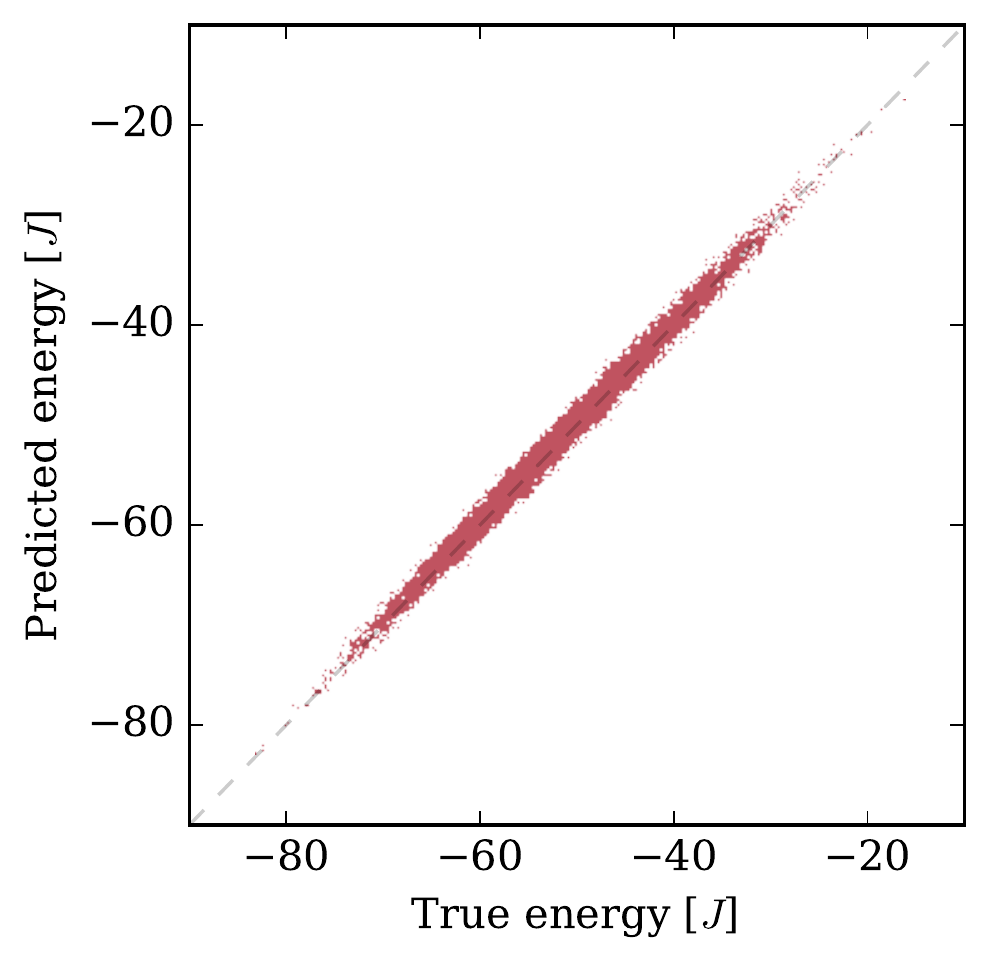}
\caption{The deep neural network is able to learn the nonbinary-state Potts Hamiltonian with a MAE of \pottsMAE on a randomly-sampled dataset.}  \label{fig:Potts}
\end{figure}

\bold{Our approach is not limited to discrete spin values}.
The planar Potts model is a generalized form of the Ising model wherein the set of possible spin states is expanded to include more than just binary spin up/down states \cite{Wu1982}. The Hamiltonian is given by 
\begin{equation} \label{eq:Potts}
\hat H = J \sum\limits_{\{i,j\}}\cos\left(\sigma_i-\sigma_j\right).
\end{equation}
In his analysis, Potts \cite{Potts1951} used discrete spin values. We train \bold{a regression model} to compute the energy of Eq. (\ref{eq:Potts}) using a continuum of spins randomly generated on the interval $[0,\pi)$. Our neural network is able to learn the mapping after observing 500,000 examples with a MAE of \pottsMAE.  In \figref{fig:Potts} we plot predicted energies against the true energies.

\subsubsection{Disordered Hamiltonians and off-lattice models}
Disordered Hamiltonians are designed to model systems where particles do not form a regular lattice.  Because of the irregular interatomic spacing, the bonds within these structures experience differing amounts of compression and expansion. ``Spin'' Hamiltonians attempt to map the disorder in atomic positions (off-lattice) to a regular (on-lattice) model with disordered interactions instead.  Treating spin-glass materials with convolutional neural networks should be possible, but encoding the disorder into the Hamiltonian (i.e. creating an on-lattice model) is not, as it introduces spatially-dependent operators, a feature inherently (and intentionally) ignored by convolutional neural networks.  Rather, one could use an off-lattice model such as those used in \cite{Ryczko2017} and \cite{Luchak2017}.

\subsubsection{DNN versus other methods}

A question one may ask is whether deep neural networks are the right tool for the job.  Certainly, there are other machine learning methods which do not involve such an expensive training process as deep neural networks demand.
In addition to a deep convolutional neural network, we tried two other commonly-used machine learning algorithms, kernel ridge regression (KRR) and random forests (RF), on various dataset sizes.  For the $8\times 8$ regression model, KRR performed at best poorly with a median absolute error of \eightKRRMAE.  RF performed much better than KRR with a MAE of \eightRFMAE (still far inferior to the deep neural network).  Additional machine learning methods have previously been demonstrated on the Ising model \cite{Portman2016}, and all present errors significantly greater than we observe with our deep neural network.

\subsubsection{A closer look at the convolutional kernels}

One might question how a convolutional neural network succeeds at learning the energies of Ising model configurations.  Convolutional neural networks optimize a set of weights, which when applied to the input in a specific way, result in an output representation of the original image that can then be interpreted by a final ``traditional'' neural network (i.e. the ``decision layer''). As such, the learned weights, or more appropriately named the kernels (convolution kernels are made up of $k\times k$ individual weights) act as ``feature detectors''.

We illustrate this through demonstration.  Consider a very simple convolutional deep neural network: a single convolutional layer with $3\times 3$ kernels, operating on a $9 \times 9$ Ising model with stride 3.  With the stride equal to the kernel size, each kernel applies to a unique region of the input space, with a given kernel only acting on each input pixel once.  We will use 512 kernels for this convolutional layer ($3\times 3\times 512$ individual weights).  The convolutional layer output is passed through a fully-connected layer of size 1024 and then reduced to a single output: the energy prediction.

\begin{figure}
 \includegraphics[width=0.999\columnwidth]{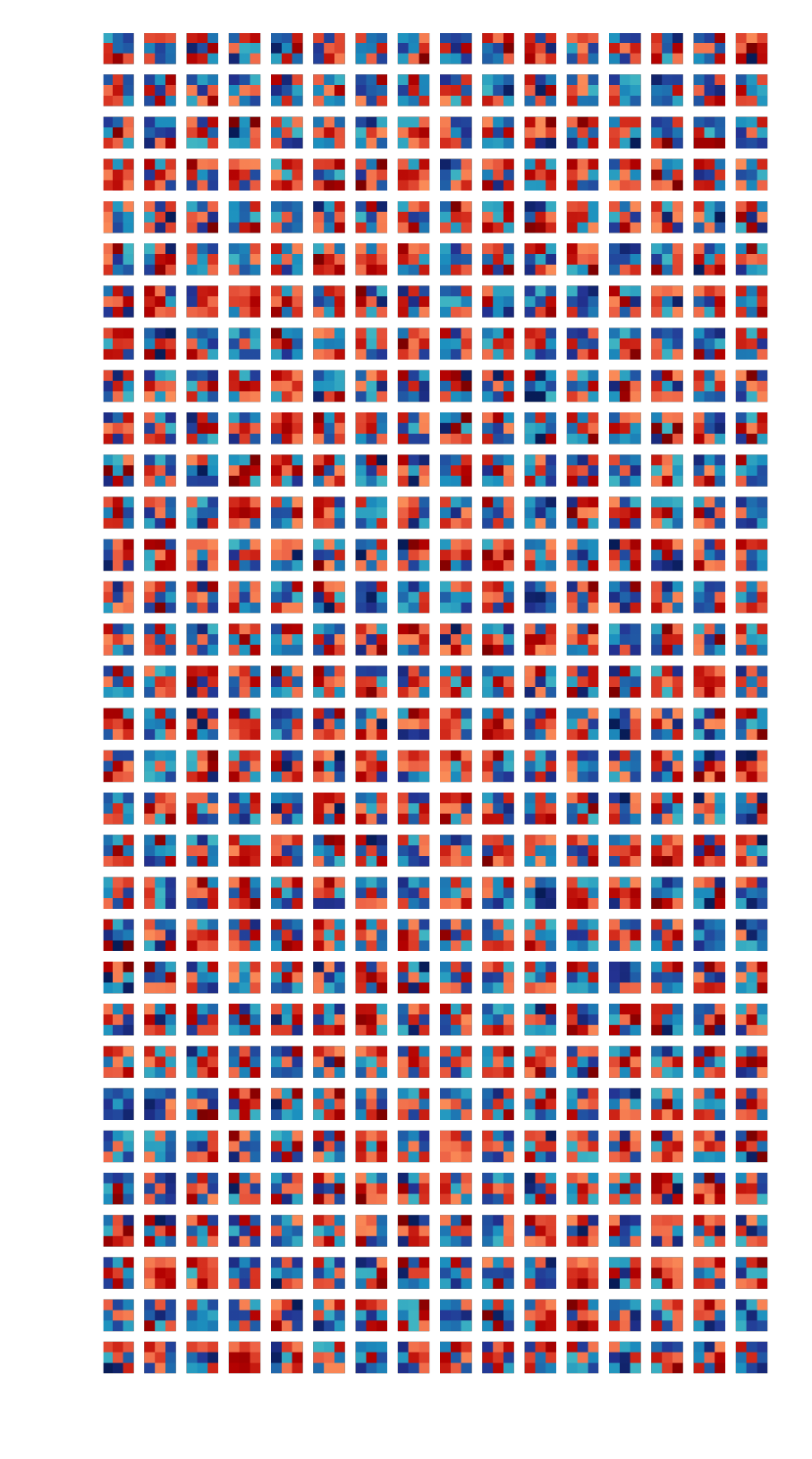}
\caption{The 512 optimized filters of our demonstration convolutional neural network. Red weights indicate negative values and blue indicate positive values.  The intensity of the color represents the magnitude of the weight.   \label{fig:512filters}}
\end{figure}

\begin{figure}
 \includegraphics[width=0.999\columnwidth]{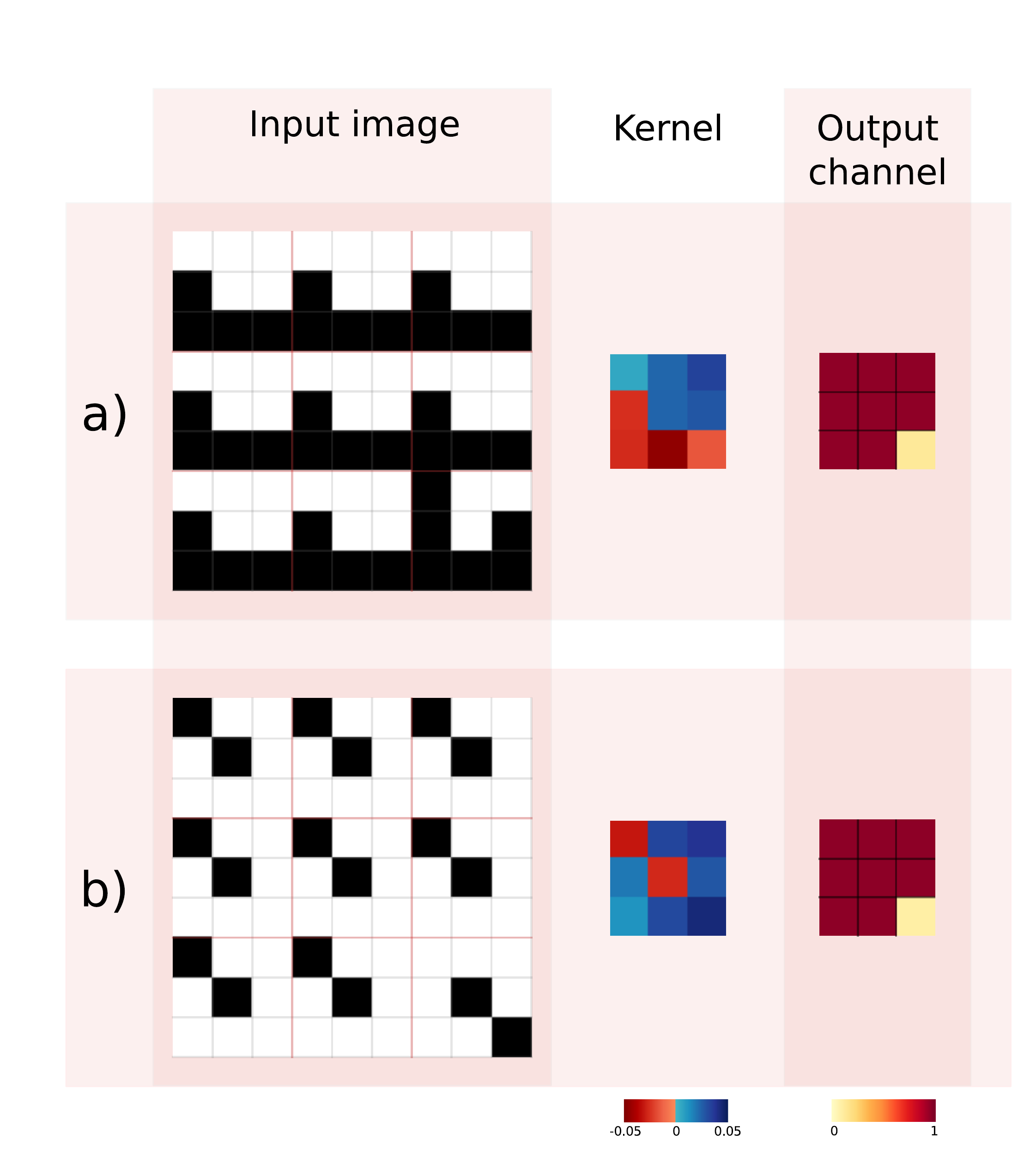}
\caption{The input images are optimized to maximize the channel output of the first (a) and last (b) filters of \figref{fig:512filters}. In both cases, we have manually set the lower right $3\times 3$ block to demonstrate how the output is affected when the filter meets a block it has not adapted to detect. \label{exampleconvolutionaloptimizationboth}}
\end{figure}

We train the neural network on 200,000 randomly generated examples, (admittedly poor practice normally, but fine for this demonstration). This simple network performs significantly more poorly than the network used throughout this work, but serves as a good example. After the network has converged, we can look at the optimized convolutional kernels.  All 512 kernels are presented in \figref{fig:512filters}.
It is difficult to tell exactly what these detect from looking at the raw weights. The magnitude of the weights are very close to zero, with some being negative (red), and some positive (blue).  We can get a better idea of what the weights have adapted to detect by finding an input image that maximizes the output of the respective channel \cite{Erhan2009}.  Using random noise as input, we can compute the gradient of the activated output with respect to the input image and optimize the image to maximize the output (gradient ascent).  This will show us the Ising configuration that maximizes the activation of the filter, and thus give us an idea of what the filter has learned to detect.  Demonstrating this on example filters produces the input images shown in \figref{exampleconvolutionaloptimizationboth}.

In this simple model, the filters learn to activate the resulting output when they see the block that they have adapted to detect. This example only works so cleanly on this very simple neural network.  In our production code, the neural network is more complex than this model, having far fewer parameters (16 or 64 per layer instead of 512), so the filters must learn to pick up on only the most relevant features, and combinations of possible features.  Additional subsequent convolutional layers provide a mechanism for the neural network to mix these features together and provide a hierarchy of feature detection, with early layers detecting more small-scale structures, and later layers picking up combinations of these small-scale features.  The final fully-connected layer then learns to take this information and map it to the energy, or magnetization. Straying from this simple example, the interpretation of the weights becomes more abstract, but the core idea remains the same:  individual kernels detect features, which when combined with all of the other kernels provide a mechanism for the neural network to map the input data to a space where interpolation is possible through the final fully-connected layers.

\subsubsection{Conclusion}

We have trained a deep neural network to accurately classify spin model configurations based on their energies. Earlier work on the Ising model has focused on the identification of phases or latent parameters through either supervised or unsupervised learning \cite{Carrasquilla2016,Wang2016a}. Following this work, we focus on learning the operators directly.  Our deep neural network learns to classify configurations based on an interacting Hamiltonian, and can use this information to make predictions about configurations it has never seen.  We demonstrate the ability of a neural network to learn the interacting Hamiltonian operator and the non-interacting magnetization operator on both the $4\times4$ and the $8\times 8$ Ising model. The performance of the larger $8\times8$ model demonstrates the ability of the model to generalize its intuition to never-before-seen examples.  We demonstrate the ability of the deep neural network in making ``physical'' predictions by replicating the phase transitions using the trained energy and magnetization operators.  In order to replicate this phase diagram, the deep neural network must use the intuition developed from observing a limited number of configurations, to evaluate configurations it has never before seen.\bold{A physical simulation such as this is the ultimate test of a mathematical model.}  Indeed it succeeds and is capable of reproducing the phase diagram precisely. We demonstrate the ability of a neural network to accurately predict the screened Coulomb interaction (a long-range interaction) \bold{and its phase transition}, and observe a speed up of three orders of magnitude over the CUDA-accelerated explicit summation. \bold{We demonstrate the ability of a deep neural network to predict the continuous-valued sinusoidally screened Coulomb Hamiltonian as well as a non-binary modified Potts Hamiltonian.}  The rapid development of featureless deep learning implementations and their ongoing successes in the technology sector motivate their consideration for physical and scientific problems.

\section{Appendix}
\subsection{Training/testing division}

When training machine learning models, it is typical to divide all available data into two sets: training and testing.  In this way, one trains the model on the training data, and then evaluates the performance of the model on the testing dataset.  It is important that the two sets are non-intersecting (i.e. no test examples appear in the training dataset) so that a fair evaluation of the generalizeability of the model is obtained.

\begin{figure}
 \includegraphics[width=0.99\columnwidth]{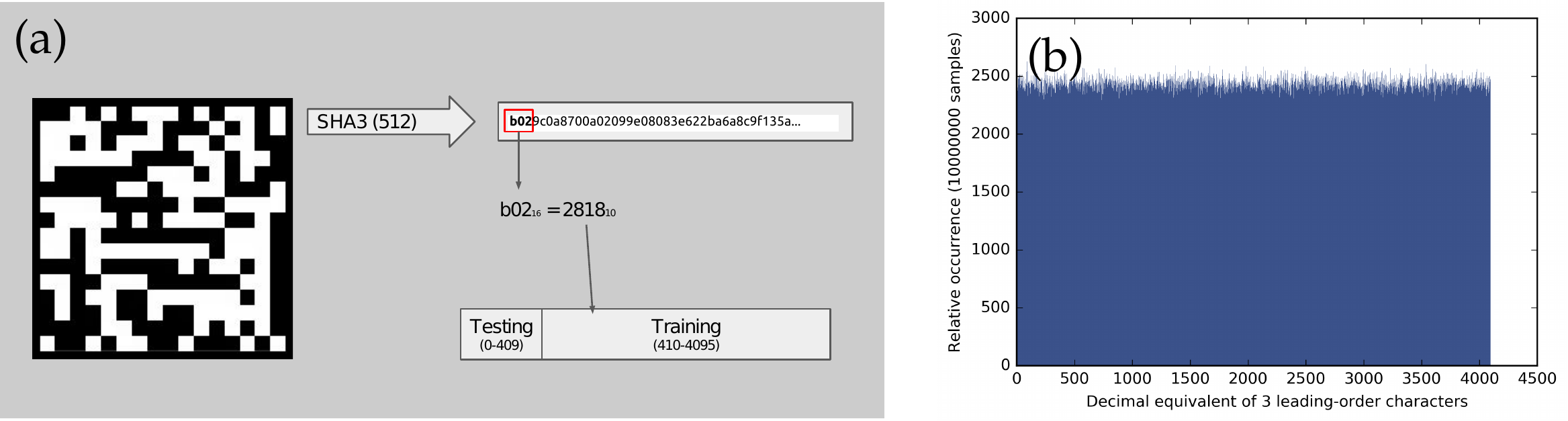}
 \caption[overview of hashing process]{(a) To randomly divide Ising configurations into training and testing datasets we compute the SHA3 hash of the configuration and use the 12 most significant bits to assign the configuration to either testing or training.  (b) This process depends on these 12 bits being uniformly distributed for many configurations.   \label{hashdescription}}
\end{figure}

In traditional machine learning applications, such as image classification, etc. this non-intersecting splitting is quite easy. Since no two images are alike, randomly assigning images to the training or testing sets is appropriate.  With the Ising model, and both sampling methods discussed above, there is the potential for duplication of training examples.  This is especially the case with targeted sampling, as duplicated training examples are necessary to achieve an even distribution of examples across the energy range. Thus, to separate examples into training and testing datasets, so that no example in the test set appears in the training set, we need  a property of the configuration that ultimately can be used to produce a binary value (e.g. 0=`test', 1=`train') in arbitrary proportions, say 10\% testing, 90\% training.  We can easily obtain a unique identifier by converting the configuration to a binary value, but the binary value is correlated to the energy, thus the split would not be random.  We could randomize a static one-to-one mapping to solve this issue, but storing such a mapping, even for the $6\times 6$ Ising model would take 275 GB of memory.  Our solution to determine whether an example should be assigned to the testing or trainings set is to compute the SHA3 hash of the configuration, and obtain the 512-bit hexadecimal digest.  Then the 3 most significant hexadecimal characters (12 most significant bits) determine whether the example gets assigned to the test set or the training set.  This allows splitting into arbitrary proportions at a resolution of 1/4096.  \figref{hashdescription} (a) shows a schematic of the process.  This procedure depends on 12 most significant bits of the SHA3 hash being uniformly distributed, and indeed they are as shown in \figref{hashdescription} (b).

\wcexclude{
\explicitsection{Acknowledgements}
The authors acknowledge funding from NSERC and SOSCIP. Compute resources were provided
by SOSCIP, National Research Council of Canada, and an NVIDIA Faculty Hardware Grant.
}

\wcexclude{
\bibliography{MSc-isingPaper}
}

\clearpage

\end{document}